\documentstyle[12pt]{article}
 \begin{document}
 \thispagestyle{empty}
 \centerline{{\large {\bf UNAVOIDABLE CONFLICT BETWEEN}}}
 \centerline{{\large{\bf MASSIVE GRAVITY
 MODELS AND}}}
 \centerline{{\large{\bf MASSIVE TOPOLOGICAL TERMS}}}

 \vskip 1.cm
 \begin{minipage}[1]{13cm}
 \centerline{ ANTONIO ACCIOLY$^*$ and MARCO DIAS}
 \centerline{{\it Instituto de F\'{\i}sica  Te\'{o}rica,
Universidade Estadual Paulista, }}
 \centerline {{\it Rua Pamplona 145, 01405-000 S\~ao Paulo, SP, Brazil}}
 \centerline {{\it $^*$accioly@ift.unesp.br}}
 \vskip .7cm
 \vskip.3cm  Massive gravity models in 2+1 dimensions, such as those
obtained by adding to Einstein's gravity the usual Fierz-Pauli, or
the more complicated Ricci scalar squared ($R^2$), terms, are tree
level unitary. Interesting enough these seemingly harmless systems
have their unitarity spoiled when they are augmented by a
Chern-Simons term. Furthermore, if the massive topological term is
added to $R + R_{\mu\nu}^2$ gravity, or to $R + R_{\mu\nu}^2 +
R^2$ gravity (higher-derivative gravity), which are nonunitary at
the tree level, the resulting models remain nonunitary. Therefore,
unlike the common belief, as well as the claims in the literature,
the coexistence between three-dimensional massive gravity models
and massive topological terms is conflicting.
 \vskip .5cm


 \end{minipage}

\newpage

 \noindent The remarkable properties of topological tensor gauge theories
in $2 + 1$ dimensions are by now not only well-appreciated but
also well-understood. The linearized versions of these models
describe single massive but gauge-invariant excitations$^1$.
Nonetheless, according to a somewhat obscure tree unitarity lore
it is expected that the operation of augmenting  a nontopological
massive gravity model through the topological term would transform
the nonunitary systems into unitary ones and preserve the tree
unitarity of the originally unitary models. In truth, this
addition does more harm than good. Indeed, innocuous avowedly tree
unitary models, such as Fierz-Pauli gravity or the more
sophisticated $R + R^2$ gravity, become nonunitary after the
topological addition, while admittedly nonunitary systems, such as
$R + R_{\mu\nu}^2$ gravity or higher-derivative gravity $(R + R^2
+ R_{\mu\nu}^2)$ remain stubbornly nonunitary after the
topological enlargement.

Our aim here is to discuss the incompatibility between massive
gravity models and massive topological terms. The analysis
comprises massive topological gravity systems that are focus of
the controversy in the literature: topological Fierz-Pauli gravity
and topological higher-derivative gravity---they are wrongly
considered as tree unitary models$^{2-4}$---and topological $R +
R^2$ and $R + R_{\mu\nu}^2$ gravity. We will show that these
topological models are without exception nonunitary at the tree
level.

To probe the tree unitarity of the models we make use of the
method that consists in saturating the propagator with external
conserved currents, $T^{\mu\nu}$, compatible with the symmetries
of the theory. The unitarity analysis is based on the residues of
the saturated propagator ($SP$): the tree unitarity is ensured if
the residue at each pole of the $SP$ is positive. Note that the
$SP$ is nothing but the current-current amplitude in momentum
space.

Natural units are used throughout. Our signature is $(+, -, -)$.
The Riemann and Ricci tensors are defined respectively as
${R^\rho}_{\lambda\mu\nu} = - \partial_\nu
{\Gamma^\rho}_{\lambda\mu} + \partial_\mu
{\Gamma^\rho}_{\lambda\nu} - {\Gamma^\sigma}_{\lambda\mu}
{\Gamma^\rho}_{\sigma\nu} + {\Gamma^\sigma}_{\lambda\nu}
{\Gamma^\rho}_{\sigma\mu} $ and $R_{\mu\nu} =
{R^\rho}_{\mu\nu\rho}.$

We consider first topological Fierz-Pauli gravity (TFPG). The
Lagrangian related to this model is the sum of Einstein, standard
Fierz-Pauli, and Chern-Simons, terms, namely,

\begin{eqnarray}
 {\cal{L}} =a \frac{2}{{\bar \kappa}^2} {\sqrt{g}} \; R - \frac{m^2}{2} \left( {h_{\mu\nu}}^2 -
h^2 \right) + \frac{1}{\mu} \epsilon^{\lambda\mu\nu}
{\Gamma^{\rho}}_{\lambda\sigma} \left( \partial_\mu
{\Gamma^\sigma}_{\rho\nu} +\frac{2}{3}
{\Gamma^\sigma}_{\mu\beta}{\Gamma^\beta}_{\nu\rho}\right),
\end{eqnarray}

\noindent at quadratic order in ${\bar \kappa}$, where $ {\bar
\kappa}^2$ is a suitable constant that in four dimensions is equal
to $24\pi G$, with $G$ being  Newton's constant$^5$. Here
$g_{\mu\nu} \equiv \eta_{\mu\nu} + {\bar \kappa} h_{\mu\nu},$ \,$h
\equiv \eta_{\mu\nu} h^{\mu\nu},$ and $a$ is a convenient
parameter that can take the values $+1$ (Einstein's term with the
usual sign)  or $-1$ (Einstein's term with the ``wrong" sign), so
that this is the most general such model. From now on indices are
raised (lowered) with $\eta^{\mu\nu}$ ($\eta_{\mu\nu}$).

 To compute the SP we have to find beforehand the propagator,
which involves much algebra. However, the calculations are great
simplified if we appeal to a set of operators made up by the usual
three-dimensional Barnes-Rivers operators$^6$, i.e.,

\begin{eqnarray}
P^1_{\mu\nu,\;\rho\sigma}=\frac{1}{2} \left( \theta_{\mu\rho}\;
\omega_{\nu\sigma} + \theta_{\mu\sigma}\;\omega_{\nu\rho} +
\theta_{\nu\rho}\;\omega_{\mu\sigma} +
\theta_{\nu\sigma}\;\omega_{\mu\rho} \right),
\nonumber
\end{eqnarray}

\begin{eqnarray}
P^2_{\mu\nu,\;\rho\sigma} = \frac{1}{2} \left(
\theta_{\mu\rho}\;\theta_{\nu\sigma} +
\theta_{\mu\sigma}\;\theta_{\nu\rho} -
\theta_{\mu\nu}\;\theta_{\rho\sigma} \right), \nonumber
\end{eqnarray}

\begin{eqnarray}
P^0_{\mu\nu,\;\rho\sigma} = \frac{1}{2}
\theta_{\mu\nu}\;\theta_{\rho\sigma},\;\;{\overline
P}^{\;0}_{\mu\nu,\;\rho\sigma} =
\omega_{\mu\nu}\;\omega_{\rho\sigma}, \nonumber
\end{eqnarray}

\begin{eqnarray}
{\overline{\overline P}}^{\;0}_{\mu\nu,\;\rho\sigma} =
\theta_{\mu\nu}\;\omega_{\rho\sigma} +
\omega_{\mu\nu}\;\theta_{\rho\sigma}, \nonumber
\end{eqnarray}

\noindent where $\theta_{\mu\nu}$ and $\omega_{\mu\nu}$ are the
well-known transverse and longitudinal vector projector operators
$\theta_{\mu\nu}= \eta_{\mu\nu} -\frac{\partial_\mu
\partial_\nu}{\Box},\;$ $\omega_{\mu\nu} =\frac{\partial_\mu
\partial_\nu}{\Box},$ and the operator

\begin{eqnarray}
P_{\mu\nu,\;\rho\sigma} \equiv \frac{\Box
\partial^\lambda}{4}[\epsilon_{\mu\lambda\rho}\;\theta_{\nu\sigma}
+ \epsilon_{\mu\lambda\sigma}\;\theta_{\nu\rho} +
\epsilon_{\nu\lambda\rho}\;\theta_{\mu\sigma} +
\epsilon_{\nu\lambda\sigma}\;\theta_{\mu\rho}], \nonumber
\end{eqnarray}

\noindent which has its origin in the linearization of the
Chern-Simons term, i.e.,

\begin{eqnarray}
{{\cal L}_{C.S.}}_{lin} = \frac{1}{2} \frac{1}{M} h^{\mu\nu}
P_{\mu\nu,\;\rho\sigma} h^{\rho\sigma}, \nonumber
\end{eqnarray}

\noindent where $M \equiv \mu/ {\bar \kappa}^2.$ The corresponding
multiplicative table is displayed in Table 1.

Adapting to $2 + 1$ dimensions, with the help of the data from
Table 1, the algorithm for calculating the propagator related to
four-dimensional gravity theories$^7$ and using the resulting
prescription, we find that the propagator for TFPG assumes the
form

\begin{eqnarray}
O^{-1} &=& -\frac{1}{m^2} P^1 - \frac{M^2 \left( m^2 + a \Box
\right)}{\Box^3 + M^2 a^2 \Box^2 + 2 a m^2 M^2 \Box + M^2 m^4} P^2
- \frac{m^2 + a \Box}{2m^4} {\overline P}^{\;0} \nonumber \\ &&+
\frac{1}{2m^2} {\overline{\overline P}}^{\;0} - \frac{M}{\Box^3 +
M^2 a^2 \Box^2 + 2am^2M^ 2 \Box + M^2m^4 } P.\nonumber
\end{eqnarray}

Consequently, the saturated propagator is given by $$ {SP}_{TFPG}
= T^{\mu\nu} O^{-1}_{\mu\nu\;,\;\rho\sigma} T^{\rho\sigma}.$$
\noindent Performing the computations, we promptly obtain in
momentum space
\begin{eqnarray}
{SP}_{TFPG}  = \left[ T^{\mu\nu}(k) T_{\mu\nu}(k) - \frac{1}{2}
T^2(k) \right] \frac{M^2 \left(m^2 -k^2a \right) }{k^6 -M^2a^2k^4
+ 2am^2M^2k^2 -M^2m^4}. \nonumber
\end{eqnarray}

At first sight it seems that we should start our analysis by
setting $a=-1$ since in the limit $m^2=0$ (1) reduces to pure
massive topological gravity (MTG)---a theory that requires $a=-1$
to be ghost-free$^{1}$. Nevertheless, a straightforward numerical
computation shows that among the roots of the equation
\begin{eqnarray}
x^3- a^2\lambda m^2x^2+2a\lambda m^4x-\lambda m^6=0,
\end{eqnarray}

\noindent where $x \equiv k^2$, $a=-1$ and $\lambda \equiv \left(
\frac{M}{m} \right)^2$, there are always two complex  roots for
any positive $\lambda$ value. Therefore, this model is unphysical
and must be rejected.

Now we shall concentrate our attention on the system with $a=+1$.
In this case the roots of (2) can be classified as \newpage
\begin{table}[!h]
\begin{center}
\begin{tabular}{|c||c|} \hline
    $\lambda <27/4$&one real root and two complex ones
   \\ \hline
  $\lambda = 27/4$ & three real roots: $x_1=x_2=4x_3 =4M^2/9$
  \\\hline
  $\lambda >27/4$& three distinct positive real roots \\ \hline
\end{tabular}
\end{center}
\end{table}

\noindent Accordingly, the viability of the theory requires
$\lambda >\frac{27}{4}$, which implies that the $SP_{TFPG}$ may be
written as $$SP_{TFPG} = X_ 1 + X_2 + X_3,$$ \noindent where

\begin{eqnarray}
X_1 =\frac{M^2 (m^2 - x_1) F(k)}{(x_1-x_2) (x_1-x_3)
(k^2-x_1)}\;,\; X_2 =\frac{M^2 (m^2 -x_2) F(k)}{(x_2-x_1)
(x_2-x_3)  (k^2-x_2)}, \nonumber
\end{eqnarray}

\begin{eqnarray}
X_3= \frac{M^2 (m^2-x_3) F(k)}{(x_3-x_1)(x_3-x_2)(k^2-x_3)}.
\nonumber
\end{eqnarray}

\noindent with $F(k) \equiv T^{\mu\nu} (k) T_{\mu\nu} (k) -
\frac{1}{2} T^2 (k).$ We are assuming without any loss of
generality that $x_1 > x_2 > x_3$.

Let us  then find a suitable basis for expanding the sources. The
class of independent vectors in momentum space ,

$$k^\mu \equiv \left(k^0, {\bf k}\right)\;,\;{\tilde{k}}^\mu
\equiv \left(k^0, -{\bf k}\right)\;,\;\varepsilon^\mu \equiv
\left(0,\;{\vec{\epsilon}}\;\right),$$

\noindent where ${\vec{ \epsilon}}$ is a unit vector orthogonal to
${\bf k}$, does the job. In this basis the symmetric current
tensor assumes the form

$$T^{\mu\nu} = A k^\mu k^\nu + B{\tilde k}^\mu {\tilde k}^\nu + C
\varepsilon^\mu \varepsilon^\nu + Dk^{(\mu} {\tilde k}^{\nu)} + E
k^{(\mu} \varepsilon^{\nu)} + F{\tilde k}^{(\mu}
\varepsilon^{\nu)}. $$

As a consequence,

\begin{eqnarray}
F(k)=\frac{1}{2}[(A-B)k^2]^2+\frac{C^2}{2}+\frac{k^2}{2}(E^2-F^2)-(A-B)k^2C.\nonumber
\end{eqnarray}
\noindent Assuming, as usual, that $T\geq 0$, we get  $C\leq 0$,
implying  $F(k)>0$. If $x_1 < m^2$,  Res $SP_{TFPG}|_{k^2=x_1}>0$,
Res $SP_{TFPG}|_{k^2=x_2}<0$ and Res $SP_{TFPG}|_{k^2=x_3}>0$.

The theory is causal and has two spin-2 physical particles of masses
equal to $x_1$ and $x_3$, respectively, and one spin-2 ghost of
mass $x_2$. On the other hand, if $x_1 > m^2$, the model is causal
as well and has at least one spin-2 ghost of mass   $x_1$. These
models are thus nonunitary at the tree level due to the presence
of the ghosts. \noindent Note that if we set $M^2=\infty$, we
recover pure Fierz-Pauli gravity (FPG). In this case $SP_{FPG}
=\frac{F(k)}{k^2-m^2}$ and Res $SP_{FPG}|_{k^2=m^2}>0$. Since the
residue of the $SP$ is positive, FPG is tree unitary---a
well-known result. Therefore, the topological term is responsible
for breaking down the tree unitarity of the harmless FPG.

We discuss in the following the claims in the literature$^{2,3}$
concerning the tree unitarity of TFPG with $a=+1$ and $\lambda >
\frac{27}{4}$. The authors of Refs. 2 and 3 simply affirm that the
model in hand is tree unitary; however, no explicit proof is
presented to give support to their statement. For clarity's sake,
we quote from Refs. 2 and 3, in this order:
 ``It can be checked that the condition $\lambda
 >\frac{27}{4}$ must be fulfilled in order that poles that
 correspond to tachyons and ghosts be suppressed from the
 spectrum."
  ``It is checked that tachyons and ghosts are excluded
 from the spectrum whenever $\lambda > \frac{27}{4}$." We have
 shown in detail that this is not true. Actually, the authors of
 these works have made a mistake as far as the analysis of the
 sign of the residues of $SP_{ TFPG}$ is concerned, which led them to conclude
 incorrectly that TFPG is unitary at the tree level.

\begin{center}
\begin{table}[t]
\caption{Multiplicative operator algebra fulfilled by $P^1$,
$P^2$, $P^0$, $\overline{P}^{\;0}$,$\overline{\overline{P}}^{\;0}$
and  $P$. Here
${P^{\theta\omega}}_{\mu\nu\;,\;\rho\sigma}\equiv\theta_{\mu\nu}\omega_{\rho\sigma}$
and
${P^{\omega\theta}}_{\mu\nu\;,\;\rho\sigma}\equiv\omega_{\mu\nu}\theta_{\rho\sigma}$.
} \vskip .5cm
\begin{center}
\begin{tabular}{|c||c|c|c|c|c|c|}\hline
&$P^1$&$P^2$&$P^0$&${\overline P}^{\;0}$&${\overline{\overline P}
}^{\;0}$&$P$\\\hline
 $P^1$&$P^1$&0&0&0&0&0\\\hline
$P^2$&0&$P^2$&0&0&0&$P$\\\hline
$P^0$&0&0&$P^0$&0&$P^{\theta\omega}$&0\\\hline $\overline
{P}^{\;0}$&0&0&0&${\overline
P}^{\;0}$&$P^{\omega\theta}$&0\\\hline
 ${\overline{\overline P}
}^{\;0}$&0&0&$P^{\omega\theta}$&$P^{\theta\omega}$&$2(P^0+\overline
{P}^{\;0})$&0\\\hline $P$&0&$P$&0&0&0&$-{\Box}^3 P^2$\\\hline

\end{tabular}
\end{center}
\end{table}
\end{center}

We consider now topological higher-derivative gravity (THDG),
whose Lagrangian has the form
\begin{eqnarray}
{\cal L} = \sqrt{-g} \left(a \frac{2R}{\kappa^2} +\frac{\alpha}{2}
R^2 +\frac{ \beta}{2}  R_{\mu\nu}^2 \right)+ \frac{1}{\mu}
\epsilon^{\lambda\mu\nu} {\Gamma^{\rho}}_{\lambda\sigma} \left(
\partial_\mu {\Gamma^\sigma}_{\rho\nu} +\frac{2}{3}
{\Gamma^\sigma}_{\mu\beta}{\Gamma^\beta}_{\nu\rho}\right),
\end{eqnarray}
\noindent where $\kappa^2$ is a constant that in four dimensions
is equal to $32\pi G$. Here $\alpha$ and $\beta$ are suitable
dimensional constants. Before going on we must answer a crucial
question: What is the use of augmenting pure three-dimensional
gravity through the quadratic terms $R^2$ and $R_{\mu\nu}^2$? The
answer is quite straightforward: The quadratic terms convert
Einstein's gravity, that is trivial from the classical viewpoint,
into a nontrivial model; furthermore, within the quantum scheme,
higher-derivative gravity (HDG), unlike pure Einstein's gravity,
has propagating degrees of freedom. In other words, the net effect
of adding quadratic or topological terms to pure Einstein's
gravity is just the same: to produce a nontrivial gravity model
with gravitons. HDG models have interesting properties of their
own:

 \vskip .3cm \noindent (i)The general solution of its
linearized version$^{8}$ great resembles,{ \it mutatis mutantis},
the four-dimensional metric of a straight $U(1)$ gauge cosmic
string in the context of linearized four-dimensional HDG$^{9}$.
 \vskip .3cm \noindent (ii)Contrary to what happens with the Newtonian
 potential---which has a logarithmic singularity at the origin and
 is unbounded at infinity---HDG's nonrelativistic potential is
 extremely well-behaved: it is finite at the origin and zero at
 infinity$^{10}$.

\vskip .3cm \noindent (iii)Both antigravity and gravitational
shielding can coexist without conflict with HDG$^{11}$.

\vskip .3cm \noindent (iv)The gravitational deflection angle
associated to HDG is always less than that related to pure
gravity$^{8}$.

\vskip .3cm Despite these nice properties, HDG possesses a ghost
pole in the tree propagator which renders it nonunitary within the
standard pertubation scheme. However, according to the already
mentioned tree unitarity lore, it is naively expected that if we
augment HDG via the Chern-Simons term we would arrive at a tree
unitary theory$^{4}$. Our main objective in what follows is to
expose the fallacy of this conjecture.

To find the propagator concerning THDG, we linearize (3) and add
to the result the gauge-fixing Lagrangian, ${\cal L}_{g.f.} =
\frac{1}{2\lambda} \left( {h_{\mu\nu}}^{,\;\nu} - h_{,\;\mu}
\right)^2$, that corresponds to the de Donder gauge. The
propagator is given by
\begin{eqnarray}
{\cal O}^{-1}&=&
\frac{2\lambda}{k^2}P^1-\frac{2M^2(2a-b\Box)}{\Box[M^2b^2\Box^2-4(abM^2-1)\Box+4M^2a^2]}P^2\nonumber\\
&&+\;\frac{1}{\Box[a+b(\frac{3}{2}+4c)\Box]}P^0+\left[
-\frac{4\lambda}{\Box}+\frac{2}{\Box[a+b(\frac{3}{2}+4c)\Box]}\right]\overline
{P}^{\;0}\nonumber\\
&&+\;\frac{1}{\Box[a+b(\frac{3}{2}+4c)\Box]}\overline{\overline{P}}^{\;0}
+\frac{4M}{\Box[M^2b^2\Box^2-4(abM^2-1)\Box+4M^2a^2]}P,\nonumber
\end{eqnarray}
\noindent where $b\equiv\frac{\beta\kappa^2}{2}$ and
$c=\frac{\alpha}{\beta}$.

In momentum space, the associated $SP$ assumes the form:
\begin{eqnarray}
SP_{THDG}&=&(T^{\mu\nu}T_{\mu\nu}-\frac{1}{2}T^2)\left[
-\frac{1+\sqrt{1-2abM^2}}{2a\sqrt{1-2abM^2}(k^2-M_1^2)}\right]\nonumber\\
&&+\;(T^{\mu\nu}T_{\mu\nu}-\frac{1}{2}T^2)\left[
-\frac{-1+\sqrt{1-2abM^2}}{2a\sqrt{1-2abM^2}(k^2-M_2^2)}\right]\nonumber\\
&&+\;\frac{T^{\mu\nu}T_{\mu\nu}-T^2}{ak^2}+\frac{\frac{1}{2}T^2}{a(k^2-m^2)},
\end{eqnarray}
\noindent where
\begin{eqnarray*}
M_1^2&\equiv&
\left(\frac{2}{b^2M^2}\right)[1-abM^2-\sqrt{1-2abM^2}],\nonumber\\
M_2^2&\equiv&
\left(\frac{2}{b^2M^2}\right)[1-abM^2+\sqrt{1-2abM^2}],\nonumber\\
m^2&\equiv& \frac{a}{b(\frac{3}{2}+4c )}.
\end{eqnarray*}

 We are now ready to analyse the excitations and mass counts for
generic signs and values of the parameters and for both allowed
signs of $a$. To begin with, we set $a=-1$. The absence of
tachyons in the dynamical field requires $b>0$ and $\frac{3}{2} +
4c<0$ or $-\frac{1}{2} < bM^2 <0$ and $\frac{3}{2} + 4c>0$. The
former leads to Res $SP_{THDG}|_{k^2=M_1^2}
>0$, Res $SP_{THDG}|_{k^2=M_2^2} >0$, Res $SP_{THDG}|_{k^2=0} =0$
and Res $SP_{THDG}|_{k^2=m^2} <0$, while the latter results in Res
$SP_{THDG}|_{k^2=M_1^2}
>0$, Res $SP_{THDG}|_{k^2=M_2^2} <0$, Res $SP_{THDG}|_{k^2=0} =0$
and Res $SP_{THDG}|_{k^2=m^2} <0$. The particle content related to
the first situation is two  massive spin-2 physical particles, one
massless spin-2 nonpropagating particle and one massive spin-0
ghost, whereas that concerning the second one is one massive
spin-2 physical particle, one massive spin-2 ghost, one
nonpropagating graviton and one massive spin-0 ghost. Therefore,
unlike the claim in the literature$^{4}$, THDG with Einstein's
term with the ``wrong" sign  is tree nonunitary. Note that the
authors of Ref. 4 state that ``the spin-0 sector displays a
massless pole along with massive poles" as well as that ``the
massive gravitons propagate as in the pure Einstein-Chern-Simons
model: negative-norm states do not appear that spoil the spectrum,
which does not affect the unitarity", pure and simple. A quick
glimpse at Table 2 it is sufficient to convince anyone of the
wrongness of these affirmations. In truth, the authors of this
work made a mistake in examining both the excitations and mass
counts of THDG with $a=-1$.

Could it be that if we have chosen $a=+1$ we would have arrived at
an unitary system? The response to this question is negative.
Indeed, assuming (i)$b<0$ and $\frac{3}{2} + 4c <0$ or (ii) $0<
bM^2 < \frac{1}{2}$ and $\frac{3}{2} + 4c >0$ in order to get rid
of the tachyions, we come to the conclusion that (i)Res
$SP_{THDG}|_{k^2=M_1^2} <0$, Res $SP_{THDG}|_{k^2=M_2^2 }<0$, Res
$SP_{THDG}|_{k^2=0} =0$ and Res $SP_{THDG}|_{k^2=m^2}
>0$, and (ii)Res $SP_{THDG}|_{k^2=M_1^2} <0$, Res
$SP_{THDG}|_{k^2=M_2^2 }>0$, Res $SP_{THDG}|_{k^2=0} =0$ and Res
$SP_{THDG}|_{k^2=m^2}
>0$, implying that the model has (i)two massive spin-2 ghosts, one
massless spin-2 nonpropagating particle and one massive spin-0
physical particle or (ii)one massive spin-2 ghost, one massive
spin-2 physical particle, one nonpropagating graviton and one
massive spin-0 physical particle. These systems, as the preceding
ones, are also nonunitary at the tree level. The above is
summarized in Table 2. The remaining systems are tachyonic.

\begin{table}[t]
\caption{Unitarity analysis of topological higher-derivative
gravity}
\begin{tiny}
\begin{center}
\begin{tabular}{|c|c|c|c|c|c|}\hline
$a$ & $b$ & $\frac{3}{2}+4c$ &\begin{tabular}{c} excitations and\\
mass counts \end{tabular}& tachyons & unitarity
\\\hline $- 1$&$>0$&$<0$& \begin{tabular}{c} 2 massive\\
spin-2 normal particles\\ 1 massless spin-2\\ nonpropagating
particle \\ 1 massive spin-0\\ ghost
\end{tabular}
&no one& \begin{tabular}{c}nonunitary\\ at the \\ tree level
\end{tabular}\\\hline$-1$ &$\frac{-1}{2M^2}<b<0$&$>0$& \begin{tabular}{c} 1 massive\\
spin-2 normal particle\\ 1 massless spin-2\\ nonpropagating
particle\\ 1 massive spin-2\\ ghost
\\ 1 massive spin-0\\ ghost
\end{tabular}
&no one& \begin{tabular}{c}nonunitary\\ at the \\ tree level
\end{tabular}\\\hline
 +1&$<0$&$<0$&
\begin{tabular}{c} 2 massive spin-2 ghosts\\ 1
massless spin-2 \\ nonpropagating particle \\ 1  massive\\ spin-0
normal particle
\end{tabular}
&no one&\begin{tabular}{c}nonunitary\\ at the\\ tree level
\end{tabular}\\\hline
$+1$ &$0<b<\frac{1}{2M^2}$&$>0$& \begin{tabular}{c} 1 massive\\
spin-2 normal particle\\ 1 massless spin-2\\ nonpropagating
particle\\ 1 massive spin-2\\ ghost
\\ 1 massive spin-0\\ normal particle
\end{tabular}
&no one& \begin{tabular}{c}nonunitary\\ at the \\ tree level
\end{tabular}\\\hline
\end{tabular}
\end{center}
\end{tiny}
\end{table}

 We discuss now the tree unitarity of the models obtained from
 THDG by judiciously chosing the parameters $\alpha,\; \beta$ and
 $M^2$ as well as the signs of $a$.

\vskip .3cm

\noindent $\bullet$ {\it Pure Massive Topological Gravity}
$(\alpha=\beta=0)$ $$SP= - \frac{T^{\mu\nu}T_{\mu\nu}- \frac{1}{2}
T^2}{a(k^2 -a^2 M^2)}+\frac{T^{\mu\nu} T_{\mu\nu} - T^2}{ak^2}$$

\noindent $a=-1$: One massive spin-2 physical particle and one
nonpropagating graviton; the model is nontachyonic and tree
unitary.

\noindent$a=+1$: One massive spin-2 ghost and one nonpropagating
graviton; the system is nontachyonic and nonunitary at the tree
level.

\noindent {\it Comment} : Pure massive topological gravity is
unitary if and only if the sign of the Einstein's term is chosen
to be $-1$. This result was obtained in Ref. 1 using a quite
different approach.

\vskip .3cm

\noindent $\bullet$ {\it Pure $R+ R^2_{\mu\nu}$ Gravity} $(\alpha=
0, M^2=\infty)$

$$SP= - \frac{T^{\mu\nu}T_{\mu\nu}- \frac{1}{2} T^2}{a(k^2
+\frac{2a}{b})}+\frac{T^{\mu\nu} T_{\mu\nu} - T^2}{ak^2} +
\frac{\frac{1}{2} T^2}{a (k^2- \frac{3}{2}b)} $$

\noindent $a=-1$ and $b>0$: One spin-2 physical particle of mass
$m_2 ={\sqrt \frac{2}{b}}$, one nonpropagating graviton and one
spin-0 ghost of mass $m_0={\sqrt \frac{3b}{2}}$; the model is
nontachyonic and tree nonunitary.

\noindent The remaining models are unphysical because they have
complex masses.

\noindent {\it Comment} : Pure $R + R^2_{\mu\nu}$ gravity
possesses no tachyons if and only if $a=-1$ (Einstein's term with
the ``wrong" sign) and $b>0$. Nonetheless, the model has a massive
scalar ghost which renders it nonunitary at the tree level.

\vskip .3cm

\noindent $\bullet$ {\it Topological $R + R^2_{\mu\nu}$ Gravity}
$(\alpha=0)$

\begin{eqnarray}
SP&=&(T^{\mu\nu}T_{\mu\nu}-\frac{1}{2}T^2)\left[
-\frac{1+\sqrt{1-2abM^2}}{2a\sqrt{1-2abM^2}(k^2-M_1^2)}\right]\nonumber\\
&&+\;(T^{\mu\nu}T_{\mu\nu}-\frac{1}{2}T^2)\left[
-\frac{-1+\sqrt{1-2abM^2}}{2a\sqrt{1-2abM^2}(k^2-M_2^2)}\right]\nonumber\\
&&+\;\frac{T^{\mu\nu}T_{\mu\nu}-T^2}{ak^2}+\frac{\frac{1}{2}T^2}{a(k^2-\frac{3}{2}
b )}\nonumber
\end{eqnarray}

\noindent $a=-1$ and $b>0$: Two spin-2 physical particles of
masses equal to $M_1$ and $M_2$, one propagating graviton and one
spin-0 ghost of mass $m_0={\sqrt \frac{3b}{2}}$; the system is
nontachionic and tree nonunitary.

\noindent $a=+1$ and $0<b< \frac{1}{2M^2}$: One spin-2 ghost of
mass $M_1$, one spin-2 physical particle of mass $M_2$, one
nonpropagating graviton and one scalar physical particle of mass
$m_0 = {\sqrt \frac{3b}{2}}$; the model is nontachyonic and tree
nonunitary.

\noindent The other models are tachyonic.

\noindent {\it Comment} : The topological term does not cure the
nonunitarity of pure $R + R^2_{\mu\nu}$ gravity.

\vskip .3cm

\noindent $\bullet$ {\it Pure $R + R^2$ Gravity}
$(\beta=0,M^2=\infty)$

\begin{eqnarray}
SP&=&\frac{T^{\mu\nu}T_{\mu\nu}-\frac{1}{2}T^2}{ak^2}+\frac{\frac{1}{2}T^2}{a(k^2-\frac{\alpha\kappa^2}{2})}
 \nonumber
\end{eqnarray}

\noindent $a=+1$ and $\alpha>0$: One scalar physical particle of
mass $m_0=\sqrt{\frac{\alpha\kappa^2}{2}}$ and one nonpropagating
graviton; the model is nontachyonic and tree unitary.

\noindent $a=-1$ and $\alpha>0$: One scalar ghost of mass
$m_0=\sqrt{\frac{\alpha\kappa^2}{2}}$ and one nonpropagating
graviton; the system is nontachyonic and tree nonunitary.

\noindent The remaining systems are tachyonic.

\noindent {\it Comment} : $R + R^2$ gravity is tree unitary if and
only if the sign of the Einstein's term is the conventional one.

\vskip .3cm

\noindent $\bullet$ {\it Topological $R + R^2$ Gravity}
$(\beta=0)$

\begin{eqnarray}
SP&=&-
\frac{T^{\mu\nu}T_{\mu\nu}-\frac{1}{2}T^2}{a(k^2-a^2M^2)}+\frac{T^{\mu\nu}T_{\mu\nu}-\frac{1}{2}T^2}{ak^2}
+\;\frac{\frac{1}{2}T^2}{a(k^2-\frac{\alpha\kappa^2}{2})}\nonumber
\end{eqnarray}

\noindent $a=-1$ and $\alpha>0$: One spin-2 physical particle of
mass M, one nonpropagating graviton and a  scalar ghost of mass
$m_0=\sqrt{\frac{\alpha\kappa^2}{2}}$;  the system is nontachyonic
and nonunitary at the tree level.

\noindent $a=+1$ and $\alpha>0$: One  spin-2 ghost of mass $M$ and
a massive scalar physical particle of mass
$m_0=\sqrt{\frac{\alpha\kappa^2}{2}}$; the model is nontachyonic
and tree nonunitary.

 \noindent The other models are nonphysical due to the complex masses.

\noindent {\it Comment} : The topological term spoils the tree
unitarity of the innocuous pure  $R + R^2$ gravity.

\vskip .3cm

\noindent $\bullet$ {\it Higher-Derivative Gravity} $(M^2=\infty)$

\begin{eqnarray}
SP&=&-
\frac{T^{\mu\nu}T_{\mu\nu}-\frac{1}{2}T^2}{a(k^2+\frac{2a}{b})}+\frac{T^{\mu\nu}T_{\mu\nu}-\frac{1}{2}T^2}{ak^2}
+\;\frac{\frac{1}{2}T^2}{a(k^2-m^2)}\nonumber
\end{eqnarray}

\noindent $a=+1$ , $b<0$ and $\frac{3}{2}+4c<0$: One spin-2 ghost
of mass $m_2=\sqrt{\frac{-2}{b}}$, one nonpropagating graviton and
one scalar physical particle of mass $m$; the system is
nontachyonic and nonunitary at the tree level.

\noindent $a=-1$, $b>0$ and $\frac{3}{2}+4c>0$: One  spin-2
physical particle of mass $m_2=\sqrt{\frac{2}{b}}$  and one scalar
ghost of mass $m$; the model is nontachyonic and tree nonunitary.

 \noindent The remaining models are tachyonic.

\noindent {\it Comment} : Higher-derivative gravity is nonunitary
for both choices of the sign of the Einstein's term .

To conclude we remark that the enlargement of massive
gravitational models via the topological term is a complete
nonsense. On the one hand, it does not cure the nonunitarity of
massive nonunitary systems (HDG, $R+R_{\mu\nu}^2$ gravity). On the
other hand, it spoils the unitarity of originally unitary masive
models (FPG, $R+R^2$ gravity).

\vskip 1.0cm

\noindent{\bf Acknowledgments}
 \vskip 1.0cm
\noindent A.A. thanks CNPq-Brazil for partial support while M.D.
is very grateful to CAPES-Brazil for full support.

\vskip 1.0cm
 \noindent {\bf References}
 \vskip .3cm
 \noindent 1. S. Deser, R. Jackiw  and S. Templeton
{\it Phys. Rev. Lett.} {\bf 48}, 975 (1982); {\it Annals of Phys.}
{\bf 140}, 372 (1982).
 \vskip .2cm
  \noindent 2. C. Pinheiro, G. Pires  and N. Tomimura
{\it Nuovo Cimento}  {\bf B111}, 1023 (1996).
 \vskip .2cm

\noindent 3. C. Pinheiro, G. Pires  and F. Rabelo de Carvalho {\it
Braz. J. Phys.} {\bf 27}, 14 (1997).
 \vskip .2cm
 \noindent 4. C. Pinheiro, G. Pires  and C. Sasaki {\it Gen. Rel. Grav.} {\bf
29}, 409 (1997).
 \vskip .2cm
 \noindent 5. A. Accioly, S. Ragusa, H. Mukai  and E. de Rey Neto  {\it Braz.
 J. Phys.} {\bf 30}, 544 (2000).
 \vskip .2cm
 \noindent 6. R. Rivers {\it Nuovo Cimento} {\bf 34}, 387 (1964); P. van Nieuwenhuizen {\it Nucl. Phys.}
  {\bf B60}, 478 (1973); K. Stelle {\it Phys. Rev.} {\bf D16}, 953 (1977); I. Antoniadis  and E. Tomboulis {\it Phys. Rev.}  {\bf D33}, 2756 (1986).
 \vskip .2cm

\noindent 7. A. Accioly, S. Ragusa , H. Mukai  and E. de Rey Neto
{\it Int. J. Theor. Phys.} {\bf 39}, 1599 (2000).
 \vskip .2cm

\noindent 8. A. Accioly, A. Azeredo  and H. Mukai   {\it Phys.
Lett.}  {\bf A279}, 169 (2001).
 \vskip .2cm

\noindent 9. B. Linet  and P. Teyssandier  {\it Class. Quantum
Grav.} {\bf 9}, 159 (1992).
 \vskip .2cm

\noindent 10. A. Accioly, H. Mukai  and A. Azeredo  {\it Class.
Quantum Grav.} {\bf 18}, L31 (2001).
 \vskip .2cm

\noindent 11. A. Accioly, H. Mukai and A. Azeredo  {\it Mod. Phys.
Lett.}  {\bf A16}, 1449 (2001).

\end{document}